\documentclass[]{aa}
\begin{document}
\title{The effects of driving time scales on heating in a coronal arcade.}
\author{T. A. Howson \inst{1} \and I. De Moortel \inst{1,2} \and L. E. Fyfe \inst{1}}
\institute{School of Mathematics and Statistics, University of St. Andrews, St. Andrews, Fife, KY16 9SS, U.K. \and Rosseland Centre for Solar Physics, University of Oslo, PO Box 1029  Blindern, NO-0315 Oslo, Norway}

\abstract{The relative importance of AC and DC heating mechanisms in maintaining the temperature of the solar corona is not well constrained.}
{Investigate the effects of the characteristic time scales of photospheric driving on the injection and dissipation of magnetic and kinetic energy within a coronal arcade.}
{We have conducted three dimensional MHD simulations of complex foot point driving imposed on a potential coronal arcade. We modified the typical time scales associated with the velocity driver to understand the efficiency of heating obtained using AC and DC drivers. We considered the implications for the injected Poynting flux and the spatial and temporal nature of the energy release in dissipative regimes.}
{For the same driver amplitude and complexity, long time scale velocity motions are able to inject a much greater Poynting flux of energy into the corona. Consequently, in non-ideal regimes, slow stressing motions result in a greater increase in plasma temperature than for wave-like driving. In dissipative simulations, Ohmic heating is found to be much more significant than viscous heating. For all drivers in our parameter space, energy dissipation is greatest close to the base of the arcade where the magnetic field strength is strongest and at separatrix surfaces, where the field connectivity changes. Across all simulations, the background field is stressed with random foot point motions (in a manner more typical of DC heating studies) and even for short time scale driving, the injected Poynting flux is large given the small amplitude flows considered. For long time scale driving, the rate of energy injection was comparable to the expected requirements in active regions. The heating rates were found to scale with the perturbed magnetic field strength and not the total field strength.}  
{For all driving time scales, loop-like structures are found to form spontaneously. Alongside recent studies which show power within the corona is dominated by low frequency motions, our results suggest that in the closed corona, DC heating is more significant than AC heating.} 
\maketitle


\section{Introduction}\label{sec:introduction}
Over the previous decades, many authors have proposed models to explain how the high temperatures of the solar corona are maintained \citep[for example, see reviews by][]{Klimchuk2006, Klimchuk2015, Erdelyi2007, Parnell2012, Reale2014}. It is widely accepted that the source of the required energy is the convective motions that exist at the solar photosphere. These surface flows are able to drive a flux of energy into the atmosphere, where ultimately it is converted into heat. However, the exact nature and location of energy release remains unclear. 

For the most part, the proposed mechanisms of energy release fall into one of two broad categories, namely, AC and DC heating models. This grouping arises in accordance with the characteristic time scales associated with the photospheric motions. In particular, if the time scales are short in comparison to the Alfv\'en travel time along a coronal loop, $\tau_A$, then the heating is considered to be AC in nature \citep[e.g. review by][]{Arregui2015}. Conversely, if the time scales are long in comparison to $\tau_A$, then the proposed model is classified as a DC heating mechanism \citep[e.g. review by][]{Wilmot-Smith2015}.

In typical coronal conditions, the magnetic and fluid Reynolds numbers are expected to be many orders of magnitude larger than unity and thus significant energy release requires the formation of small scales in either the magnetic or velocity fields, or indeed in both. DC heating models generally propose that the slow stressing of magnetic foot points induces the formation of intricate current sheets within the coronal volume and, in the case of finite magnetic diffusivity, inevitably leads to the dissipation of energy \citep{Parker1972, Parker1988}. 

Over recent years, increasingly sophisticated numerical modelling has allowed authors to investigate the effects of a variety of imposed boundary drivers on the release of energy within the corona. These include sequences of shearing motions \citep[e.g.][]{VanBalle1988, Galsgaard1996, Bowness2013} injecting magnetic twist through rotational driving \citep[e.g.][]{DeMoortel2006, Rappazzo2013, Reid2018, Knizhnik2019} and the more realistic excitation of the corona through convective-like flows in the lower atmosphere \citep[e.g.][]{Hardi2004, Gudiksen2005, Bingert2013, Charalambos2017}. In many such simulations, authors have identified the propensity for the system to reach a stochastic, but statistically steady state where energy is ultimately dissipated at a relatively constant rate \citep[e.g.][]{Dahlburg2012}. Additionally, \citet{Ritchie2016}, found that coherent motions will lead to large but infrequent heating events whilst more complex motions will lead to low-energy but frequent heating.

AC heating models also require a cascade of energy to small length scales in order to generate significant temperature increases. A variety of models have been proposed to increase the dissipation rate of wave energy. These include resonant absorption \citep{Ionson1978}, phase mixing \citep{Heyvaerts1983} and the formation of MHD turbulence \citep[e.g.][]{VanBalle2011, Magyar2017}. The latter phenomenon is able to develop due to the non-linear interaction of counter-propagating waves or through the development of dynamic instabilities \citep[e.g.][]{Browning1984, Terradas2008a}.

In general, each of these processes requires the presence of a non-uniform profile in the local Alfv\'en speed. This is often associated with a pre-defined density profile \citep[e.g.][]{Ruderman2002}, which has not been included within the simulations presented in this article. Despite this, resonant absorption and phase mixing are able to proceed in the absence of density structuring if the magnetic field strength and/or field line lengths are non-uniform \citep[e.g.][]{Wright1994, Wright2016, Howson2019}. In recent years, large scale, three-dimensional MHD simulations have allowed increasingly complex AC heating models to be developed. These have investigated wave energy dissipation in a variety of general settings, including in multi-threaded coronal loops \citep[e.g][]{Luna2010, Guo2019}, complex magnetic field geometries \citep{Howson2019a, Howson2020} and a stratified atmosphere which considers the connection between the corona and the dense chromosphere below \citep{Riedl2019, VanDamme2020}. 

Despite significant progress in observing capability over recent decades, direct observations of either DC or AC heating remain elusive. In terms of the former mechanism, a significant limitation is the lack of ability to directly measure the coronal magnetic field. Whilst some estimates can be obtained using seismological techniques \citep[see review by][]{DeMoortel2012a} or through field extrapolation \citep[see][for example]{Weigelmann2012}, the complexity and spatial structure of the coronal magnetic field remains poorly constrained. Thus, only indirect evidence for certain heating profiles can be obtained by comparing observations to synthetic emission data derived from simulations of energy release \citep[e.g.][]{Lionello2013, Winebarger2018}.

For AC heating, on the other hand, several studies have attempted to estimate the energy associated with coronal waves \citep[e.g.][]{Tomczyk2007, McIntosh2011}. In \citet{Morton2016}, the authors established the spectral slope of wave power in active regions, the Quiet Sun and open field regions. Each region exhibited enhanced power at approximately 3 mHz, leading the authors to posit that transverse waves are ultimately driven throughout the corona by $p$-modes in the solar interior \citep{Morton2019}. Additionally, and importantly for the current study, in all cases, the oscillatory power in low frequencies dominates over higher frequency motions. Indeed, most power is present below the expected natural Alfv\'en frequency of a typical coronal loop, suggesting, according to the classical definition, there is a much greater energy budget available for DC heating than for AC heating.

In this article, we compare the efficiency of plasma heating generated by velocity drivers with different characteristic time scales. We present the results of 3-D numerical MHD simulations of transverse motions imposed at the foot points of potential magnetic fields. We investigate the flux of energy through the numerical domain and explore the spatial distribution of heating. In Section \ref{Sec_NM}, we outline our model and discuss the nature of the imposed velocity driver. In Section \ref{Sec_Res} we present our results and in Section \ref{Sec_Dis} we discuss the implications of this study in the context of the coronal heating problem.

\section{Numerical method}\label{Sec_NM}
To obtain the results presented within this article, we have used the Lagrangian-Remap code, Lare3d \citep{Arber2001}. The scheme advances the full, three dimensional, MHD equations in normalised form, given by
\begin{equation}\frac{\text{D}\rho}{\text{D}t} = -\rho \vec{\nabla} \cdot \vec{v}, \end{equation}
\begin{equation} \label{eq:motion} \rho \frac{{\text{D}\vec{v}}}{{\text{D}t}} = \vec{j} \times \vec{B} - \vec{\nabla} P + \vec{F}_{\text{visc.}}, \end{equation}
\begin{equation} \label{eq:energy} \rho \frac{{\text{D}\epsilon}}{{\text{D}t}} = \eta \lvert \vec{j}\rvert^2- P(\vec{\nabla} \cdot \vec{v}) + Q_{\text{visc.}}, \end{equation}
\begin{equation}\label{eq:induction}\frac{\text{D}\vec{B}}{\text{D}t}=\left(\vec{B} \cdot \vec{\nabla}\right)\vec{v} - \left(\vec{\nabla} \cdot \vec{v} \right) \vec{B} - \vec{\nabla} \times \left(\eta \vec{\nabla} \times \vec{B}\right), \end{equation}
where all variables have their usual meanings. We include the resistivity, $\eta$ and viscosity, $\nu$ as non-ideal terms which dissipate energy from the magnetic and velocity fields, respectively. The viscosity is a sum of contributions from a background viscosity and two small shock viscosity terms which are included within all following simulations to ensure numerical stability. Together, these contribute a force, $\vec{F}_{\text{visc.}}$ on the right-hand side of the equation of motion (\ref{eq:motion}) and a heating term, $Q_{\text{visc}}$ to the energy equation (\ref{eq:energy}). The scheme employed here does not force energy conservation. In particular, numerical dissipation will not lead to an increase in the plasma temperature.

\subsection{Initial conditions}
\begin{figure}[h]
  \centering
  \includegraphics[width=0.45\textwidth]{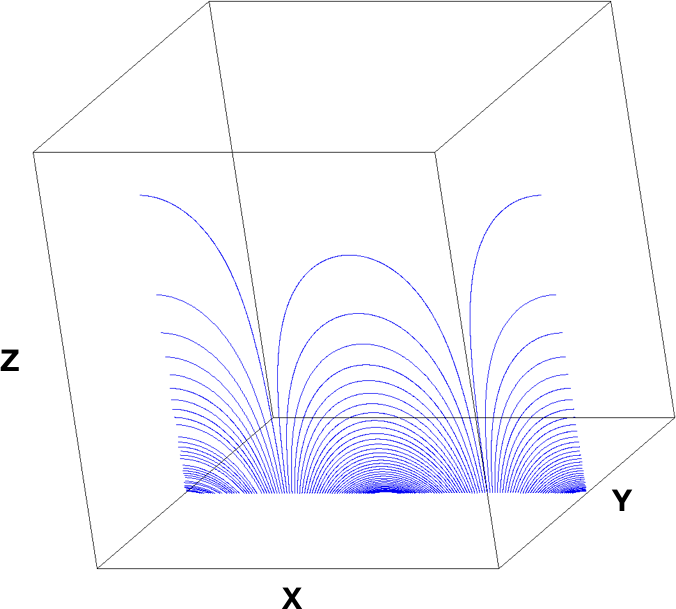}
  \caption{Magnetic field lines for the coronal arcade. The arcade is invariant in the $y$ direction.}
  \label{In_Fielda}
\end{figure}

We considered a potential coronal arcade with uniform density and temperature and we neglected the effects of gravity, thermal conduction and radiative losses. We implemented a computational domain with $256^3$ grid points and dimensions, $-L < x < L, -L < y < L$ and $0 < z < 2L$, such that the lower $z$ boundary represented the base of the corona. The magnetic field was invariant in the $y$ direction and was defined as $\vec{B} = \left(B_x, 0, B_z\right)$, where
\begin{eqnarray}
B_x(x, z) =&B_0 \cos\left(\frac{\pi x}{L}\right)\exp\left(\frac{-\pi z}{L}\right),\\
B_z(x, z) =&- B_0 \sin\left(\frac{\pi x}{L}\right)\exp\left(\frac{-\pi z}{L}\right).
\end{eqnarray} 
This field is potential and therefore force-free. Additionally, since $\nabla P = 0$, the initial conditions were in equilibrium. The nature of the magnetic field is displayed in Fig. \ref{In_Fielda}. For the following simulations, we selected $L = 10$ Mm and $B_0 = 100$ G. The initial field strength decreases exponentially with height from this value at the base of the domain to 0.2 G at the upper boundary. The initial temperature was approximately 1 MK and the initial density was $\rho_0 = 1.67 \times 10^{-12} \text{ kg m}^{-3}$. At $z=1$ Mm (base of the resistive volume; see Sect. 3.2), the plasma-$\beta$ is approximately $10^{-3}$ and the Alfv\'en speed is approximately $5000 \text{ km s}^{-1}$.

\subsection{Boundary conditions}
We seek to mimic the convective flows that exist at the photosphere by imposing a transverse, space- and time-dependent velocity profile at the base of the coronal domain. We note that our simulations do not include the chromosphere or the transition region and the precise mechanisms by which energy is transferred through these complex regions remain unclear. For the purpose of the current study, we simply assume that some of the energy from photospheric motions is transmitted to the foot points of coronal loops and that the spatial and temporal scales of the flows are similar.

We aim to replicate the turbulent-like nature of the convective flows by imposing a boundary driver that varies randomly in both space and time. We define the driver using a sum of many individual 2-D Gaussians, each of which has a particular amplitude, direction, length scale and switches on and off in time. In particular, on the lower ($z = 0$) boundary, we impose $\vec{v} = (v_x, v_y, 0)$, where
\begin{eqnarray}
\label{vx_def} v_x = \displaystyle \sum_{i=1}^N v_i \cos{\theta_i} \exp\left\{\frac{-(r-r_i)^2}{l_i^2}\right\}\exp\left\{\frac{-(t-t_i)^2}{\tau_i^2}\right\},\\
\label{vy_def} v_y = \displaystyle \sum_{i=1}^N v_i \sin{\theta_i} \exp\left\{\frac{-(r-r_i)^2}{l_i^2}\right\}\exp\left\{\frac{-(t-t_i)^2}{\tau_i^2}\right\}.
\end{eqnarray}
Here, for each $i$ in the summation, $v_i$ is the amplitude of each component, $\theta_i$ defines the direction of the driver, $r_i$ is the centre of a 2-D Gaussian, $l_i$ is a parameter which defines the length scales of the velocity driver, $t_i$ is the time of peak amplitude for each component and $\tau_i$ is used to define the lifetime of each Gaussian. 

For each term in the summation, all parameters are randomly selected from some distribution. In particular, for all $i$, the $v_i$ are normally distributed with mean $v_{\mu}$ and variance $v^2_{\mu}/25$, the $\theta_i$ are uniformly distributed on the interval $\left[0, 2 \pi\right]$, the $r_i$ are uniformly distributed over the lower boundary of the domain, the $l_i$ are normally distributed with mean $L/4 = 2.5$ Mm and variance $L^2/400 = 0.25 \text{ Mm}^2$, the $t_i$ are uniformly distributed over the duration of the simulation, and the $\tau_i$ are normally distributed with mean $\tau_{\mu}$ and variance $\tau^2_{\mu}/16$. 

For this article, we conducted a parameter study on the characteristic velocity time scale, $\tau_{\mu}$. We note that smaller values of $\tau_{\mu}$ create shorter time scales for the velocity driver. Therefore, in terms of the classical division of coronal heating models, any resultant energy dissipation will be more similar to that caused by AC heating mechanisms. Conversely, larger values of $\tau_{\mu}$ will cause slower stressing of the coronal field and be more comparable to DC heating mechanisms.

\begin{figure}[h]
  \centering
  \includegraphics[width=0.45\textwidth]{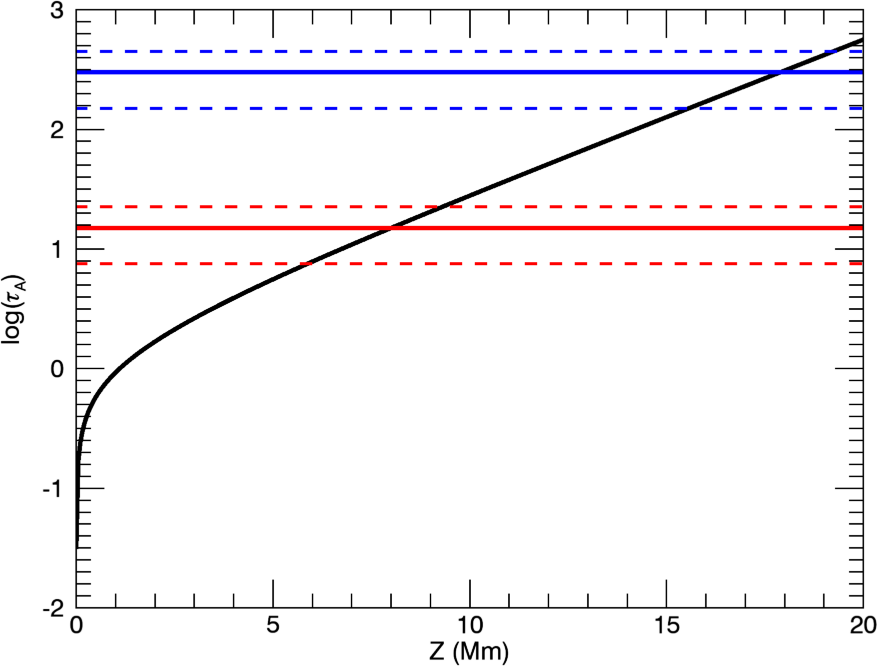}
  \caption{\emph{Black line}: Alfv\'en travel time along field lines located on the line $x=y=0$. \emph{Red and blue lines}: Mean velocity driver time scales (solid lines) for the T1 (red) and T3 (blue) simulations and the location of two standard deviations from the mean (dashed lines).}
  \label{driving_tscales}
\end{figure}

We considered simulations with three different characteristic driving time scales defined using $\tau_{\mu} \approx 15, 30$ and 300 s. Thus, using the definition from the previous section, we implement variances of 225/16, 900/16 and 90000/16 $\text{s}^2$, respectively. Henceforth, we shall refer to these simulations as $T1, T2$ and $T3$. In Fig. \ref{driving_tscales}, we display the characteristic time scales for the T1 (red line) and T3 (blue line) simulations in comparison to the Alfv\'en travel time, $\tau_A$, along magnetic field lines (black line). For clarity, we show the logarithms of these quantities. In order to find the travel time, we calculated
\begin{equation}
\tau_A(z) = \int_L \frac{\mathrm{d}s}{v_A}
\end{equation}
as a function of $z$ along the line $x=y=0$. Here, the integral is evaluated along each magnetic field line, $\mathrm{d}s$ is an infinitessimal length along the field line and $v_A$ is the local Alfv\'en speed. This produces the black curve in Fig. \ref{driving_tscales}. We note that the Alfv\'en travel time quickly converges to 0 for small values of $z$ because, close to the lower boundary, the field strength (and local Alfv\'en speed) is high and the field lines are very short. The dashed red and blue lines show the extent of two standard deviations from the mean for each of the characteristic time scales. As such, since the values $\tau_i$ (see equations \ref{vx_def} \& \ref{vy_def}) are randomly sampled from a Normal distribution, we expect approximately 95\% of the $\tau_i$ to lie within the dashed lines for the respective simulations.

\begin{figure}[h]
  \centering
  \includegraphics[width=0.5\textwidth]{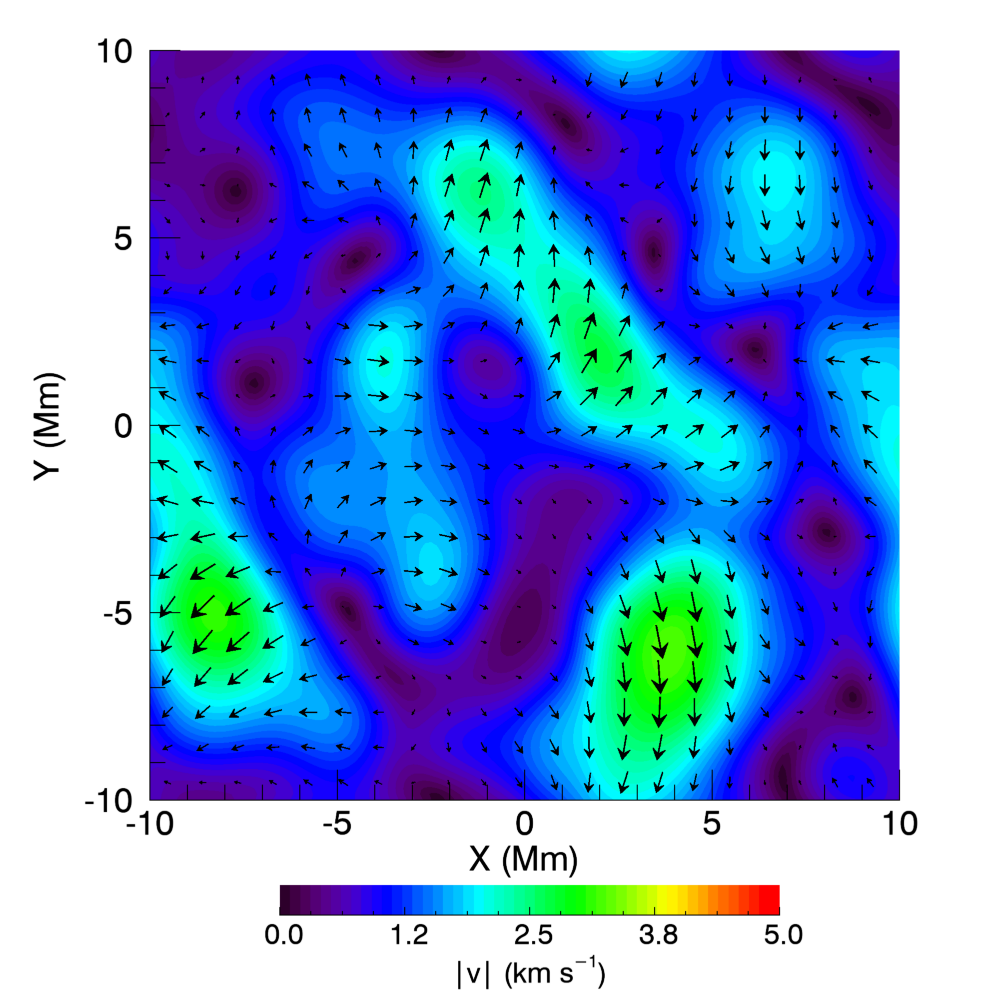}
  \caption{Contour and vector plot of the driving velocity imposed at the lower driver. We show a time of $t \approx 1800$ s for the long time scale (T3) simulation. Movies of the drivers imposed in the T1 and T3 simulations are included in the files that accompany this article.}
  \label{Driver_Example}
\end{figure}

\begin{figure}[h]
  \centering
  \includegraphics[width=0.45\textwidth]{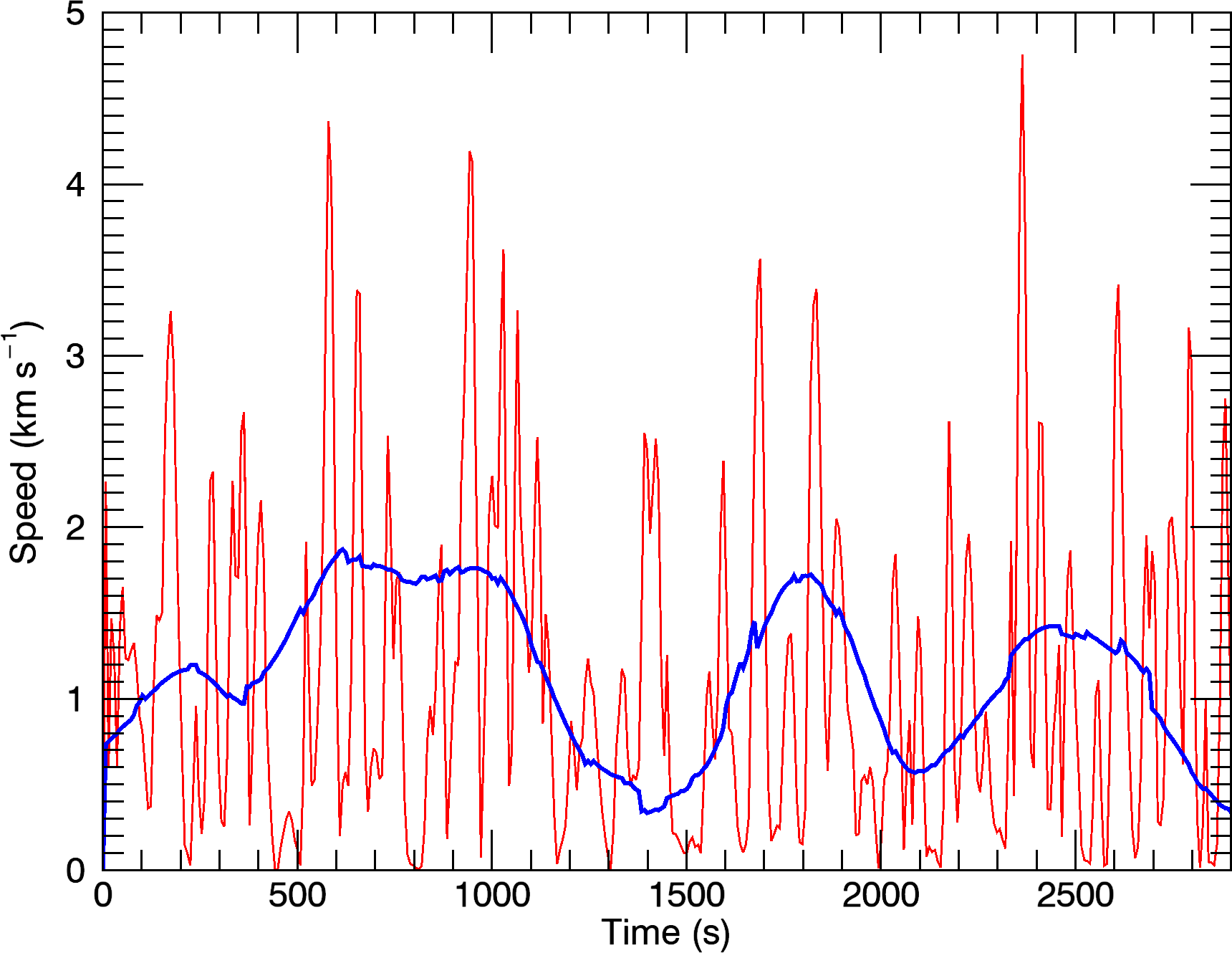}
  \caption{Imposed velocity at the centre ($x=0, y=0$) of the lower boundary for the T1 (red) and T3 (blue) simulations.}
  \label{vx_vy_point}
\end{figure}

In equations (\ref{vx_def}) \& (\ref{vy_def}), $N$, is selected to be a function of the typical time scale $\tau_{\mu}$ and is chosen such that at any time within the simulation, a similar number of components in the sum are active. This ensures that the spatial scales of the velocity driver are consistent between different simulations in the parameter space. In Fig. \ref{Driver_Example}, we show an example of the imposed driver at one instance in time for simulation T3. Movies showing the evolution of the imposed velocity field are included in the accompanying files. In Fig. \ref{vx_vy_point}, we show the time evolution of the driver in the centre of the lower $z$ boundary ($x=y=z=0$ Mm) for the T1 (red) and T3 (blue) simulations with different driving time scales. In all simulations, we selected $v_{\mu}$ such that the temporally and spatially averaged mean of the imposed velocity is approximately $1.2 \text{ km s}^{-1}$.

In all simulations, the $x$ and $y$ boundaries were set to be periodic. With the exception of the imposed velocity driver described above, a zero-gradient condition is enforced at the lower $z$ boundary for all variables. A damping layer is employed close to the upper $z$ boundary to ensure any upflows do not reflect back into the arcade. Within this layer, velocities are damped using
\begin{equation} \vec{v} = a(z) \vec{v},
\end{equation}
at every time step. Here, $a(z)$ is a damping coefficient that is equal to 1 if $z \le 18$ Mm and then decreases linearly to a value of 0.99 at $z = 20$ Mm. Despite this apparently weak damping, velocities decrease rapidly as each time step is very short in comparison to the travel time across this layer. The associated kinetic energy is simply removed from the domain and is not dissipated as heat. In reality, it would be lost to the upper corona and the solar wind which are not included within this model. At the top of the damping layer, the upper $z$ boundary enforces a zero gradient condition in all variables with the exception of velocities which are set to zero. As such, no energy flux is permitted through this boundary and any expansion of the magnetic arcades is confined to the numerical domain.


\begin{figure*}[h] \label{Isosurfaces}
  \centering
  \includegraphics[width=\textwidth]{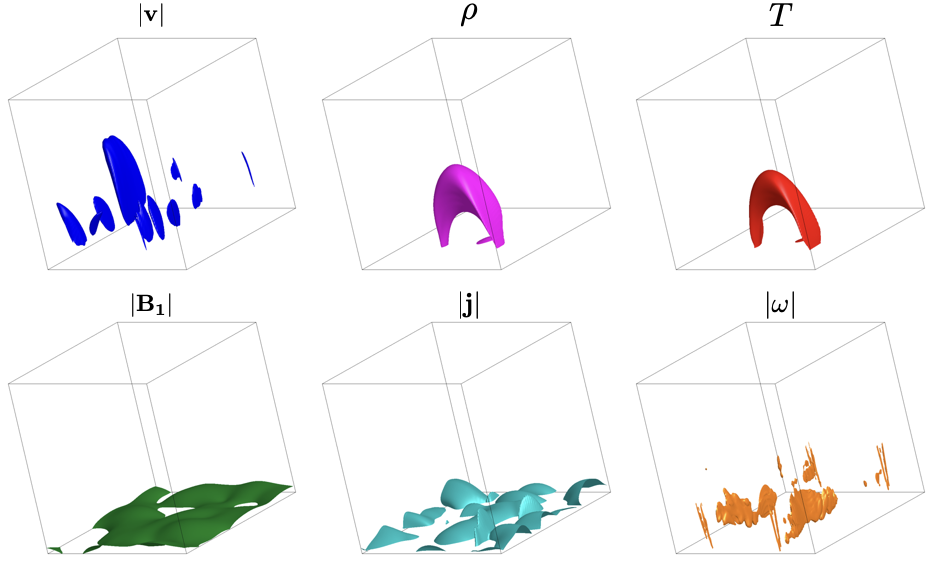}
  \caption{Isosurfaces of six variables at the end point of the T2 simulation. Clockwise from top-left: magnitude of the velocity, $|\vec{v}| = 13 \text{ km s}^{-1}$; density, $\rho = 1.5 \, \rho_0 = 2.5 \text{ kg m}^{-3}$; temperature, $T = 1.3 \, T_0 = 1.3$ MK; magnitude of vorticity, $|\vec{\omega}| = 7.6 \times 10^{-2} \text{ s}^{-1}$; magnitude of the current density, $|\vec{j}| = 8.0 \times 10^{-4} \text{ A m}^{-2}$ ; magnitude of perturbed magnetic field, $|\vec{B_1}| = 20$ G.  }
  \label{Isosurfaces}
\end{figure*}

\section{Results}\label{Sec_Res}
We begin by considering the results of the T2 simulation. At this stage, we do not include the effects of resistivity or the background viscosity (these are considered in Sect. \ref{section:non-ideal}). However, as discussed in the previous section, shock viscosities are included to ensure numerical stability. These inevitably lead to some weak, irreversible, plasma heating.

Due to the random and complex nature of the imposed velocity profile, it is impossible to account for all of the dynamics observed within each simulation. Instead, we will mostly focus on globally integrated quantities to provide comparisons between the simulations with different characteristic driving timescales.

The velocity profile imposed at the lower $z$ boundary acts to inject energy into the computational volume. The magnetic field becomes stressed and Lorentz forces drive flows throughout the domain. Velocities that are generated above $z = 0.9\, z_{\max} = 18$ Mm are rapidly reduced by the damping layer. 

In Fig. \ref{Isosurfaces}, we display isosurfaces of quantities at the end of the simulation run time. The top left-hand panel shows an isosurface of the velocity magnitude with a level of 13 $\text{km s}^{-1}$. Grid points with larger velocities are contained within the closed surfaces. We note that the largest velocities that form during the simulation are greater than those imposed at the driven boundary (see Fig. \ref{vx_vy_point}). The distribution of the isosurfaces shows that the largest velocities tend to form along separatrix surfaces at the interfaces between the two magnetic arcades ($x \approx \pm 5$ Mm). They are accelerated by large Lorentz forces that are in turn associated with the interaction of the two arcade structures. If, for example, the magnetic pressure of both arcades is increased along a fixed value of $y$, then each will exert an increased expansion force on its neighbour. Ultimately, the plasma and field are compressed, leading to large forces and the generation of flows at the arcade interface.

The remaining two panels on the top row of Fig. \ref{Isosurfaces} show isosurfaces of the plasma density (central panel) and temperature (right-hand panel). We show a level of $1.5 \, \rho_0 =  2.5 \times 10^{-12} \text{ kg m}^{-3}$ for the density and $1.3 \, T_0 = 1.3$ MK for the temperature. Again, larger values of the respective quantities are contained within the two isosurfaces. In this ideal simulation, the highest values of these two variables form co-spatially and show the spontaneous formation of a loop-like structure.  The random velocity motions sometimes induce twist within a magnetic flux tube. This induces a radially inwards tension force which will reduce the cross-section of the flux tube until it is balanced by an enhanced magnetic and (to a lesser extent) gas pressure. This increases the density and, as a result of adiabatic effects, the temperature of the plasma. Gas pressure forces are also able to accelerate field-aligned flows to spread the hotter, denser plasma along the flux tube. 

As a result of this process, the form of a magnetic feature can become identifiable in the temperature and density distributions, and, if an appropriate emission line was selected, such a structure would certainly be apparent in synthetic observables generated using the simulation data. However, in reality, with a full thermodynamic treatment of the entire solar atmosphere, the density enhancement could not be sustained against the effects of gravity, unless it was supported by an increased heating rate.

In the lower left-hand panel of Fig. \ref{Isosurfaces}, we show an isosurface of the perturbed magnetic field strength. This is calculated by subtracting the initial magnetic field from the final state. We choose only to show the perturbed field as, at all times during the simulation, the total field strength is dominated by the initial field. This means that throughout the experiments, field lines retain the approximate form of the initial arcade. We see that the largest values of the perturbed magnetic field strength are located close to the lower $z$ boundary. However, we note that this is not simply due to the proximity of the imposed driver. Instead, to understand this behaviour, we can consider the magnetic induction equation (\ref{eq:induction}) with $\eta = 0$. Since $\vec{B}$ is dominated by the background field ($\vec{B}_0$), the size of the perturbed field is governed by gradients in $\vec{v}$ and $\vec{B}_0$. These tend to be largest where $|\vec{v}|$ and $|\vec{B_0}|$ are greatest. As the magnitude of the time-averaged velocity does not strongly depend on height (see Sect. \ref{section:vel_ke}), the $z$-dependence of the perturbed field strength is related to $|\vec{B}_0|$. As such, the perturbed field strength is greatest close to the lower boundary as this is the region of largest initial field strength.

In the central panel of the second row of Fig. \ref{Isosurfaces}, we show an isosurface of the magnitude of the current density. We show a value of $8.0 \times 10^{-4} \text{ A m}^{-2}$. Since, the initial field is potential, and therefore current-free, the current density is only associated with gradients in the perturbed field. It is therefore intuitive that the largest currents form co-spatially with the highest values of the perturbed field, close to the foot points of magnetic field lines. In a resistive regime, Ohmic heating is proportional to the square of this variable and thus we would expect energy dissipation to occur most readily at low altitudes within the arcade. Whilst $|\vec{j}|$ decreases rapidly with height, at a given value of $z$, the largest currents form preferentially at the separatrix surfaces between the coronal arcades. A similar effect is observed in the magnitude of the vorticity (next paragraph).

In the final panel of Fig. \ref{Isosurfaces}, we show an isosurface of the magnitude of the vorticity, a measure of small scales in the velocity field. We show a level of $7.6 \times 10^{-2} \text{ s}^{-1}$ and note that the largest vorticities are contained within these surfaces. We see that the largest velocity gradients form along separatrix surfaces between the coronal arcades. As with the spatial correlation between the perturbed magnetic field and the current density, this is largely due to the largest velocities forming here (see top-left panel of Fig. \ref{Isosurfaces}). Additionally, the change in connectivity across the magnetic boundary plays a role in the large gradients that form. Flows excited on a field line on one side of the separatrix surface may not be present on the other side as the field line foot points may not be close together. In any regime where energy dissipation is dominated by viscous effects, we would expect plasma heating to be greatest in the regions of greatest vorticity. 

\subsection{Energy flux}
With the exception of the damping layer near the top of the domain, flows into and out of the numerical grid are not permitted, Since the damping layer reduces flows, the only energy injection is associated with a Poynting flux through the lower $z$ boundary. In an ideal regime, the change in the volume integrated energy can be expressed as 
\begin{equation} \frac{1}{\mu} \int \vec{E} \times \vec{B} \, \cdot \, \mathrm{d}\vec{S} =  \frac{1}{\mu}  \int \left\{\left(\vec{B} \cdot \vec{v}\right)\vec{B} - \left(\vec{B}\cdot \vec{B}\right)\vec{v}\right\} \, \cdot \, \mathrm{d}\vec{S}, \end{equation}
where $\vec{E}$ is the electric field and the integral is computed over the lower boundary. Since $v_z$ is set to 0 on this plane, the second term in the integrand provides no contribution and the energy flux reduces to
\begin{equation} \label{Poynt_flux_exp}\frac{1}{\mu} \int \vec{E} \times \vec{B} \, \cdot \, \mathrm{d}\vec{S} =  \frac{-1}{\mu}\int \int \left(v_xB_x + v_yB_y\right)B_z \,\, \mathrm{d}x\mathrm{d}y. 
\end{equation}
\begin{figure}[h]
  \centering
  \includegraphics[width=0.47\textwidth]{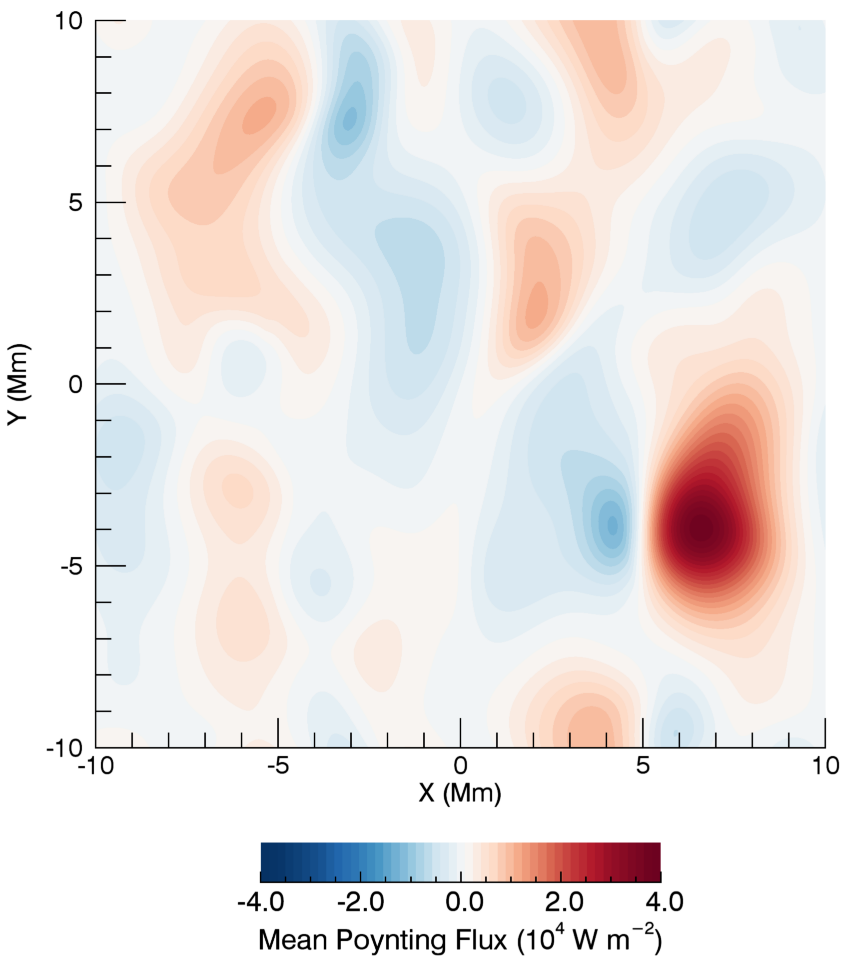}
  \caption{Mean Poynting flux through each point on the driven boundary during the ideal T2 simulation.}
  \label{Poynting_contour}
\end{figure}

In Fig. \ref{Poynting_contour}, we show the time-averaged Poynting flux through the lower boundary of the ideal T2 simulation. Red colours identify locations where the net flux of energy is into the domain and blue colours show points where energy is, on average, lost from the domain. The peak energy inflows (approximately $4 \times 10^4 \text{ Wm}^{-2})$ are in excess of the requirements for heating active regions \citep[$10^4 \text{ Wm}^{-2}$][]{Withbroe1977}, however, significant energy is also lost through the boundary. In particular, the driver is able to remove energy from both the initial and perturbed magnetic fields. Although the initial field is potential (and therefore representative of the minimum magnetic energy state for a given set of boundary conditions), magnetic energy can still be removed from the domain by modifying the distribution of magnetic flux through the boundary. Further, the driver is also able to interact with the perturbed magnetic field to remove energy from the domain. Thus, although it is still positive over the course of the simulation, the net Poynting flux (see below) is actually much smaller than the maximum values in Fig. \ref{Poynting_contour}.

Returning to equation \ref{Poynt_flux_exp}, we see that under the assumption that the field does not evolve from the initial state, the $v_xB_xB_z$ term will dominate as $B_y$ is initially zero. Whilst this assumption is not completely valid throughout the simulation, it does suggest (particularly for early times) that the magnitude of the Poynting flux will be greatest in locations where the product $B_xB_z$ is largest in the initial conditions. Of course, the energy flux is also heavily dependent on the random spatial distribution of the imposed velocity. However, since this is distributed uniformly in space, there is no systematic preference for energy flux that arises from the $v_x$ and $v_y$ terms in equation \ref{Poynt_flux_exp}.
\begin{figure}[h]
  \centering
  \includegraphics[width=0.5\textwidth]{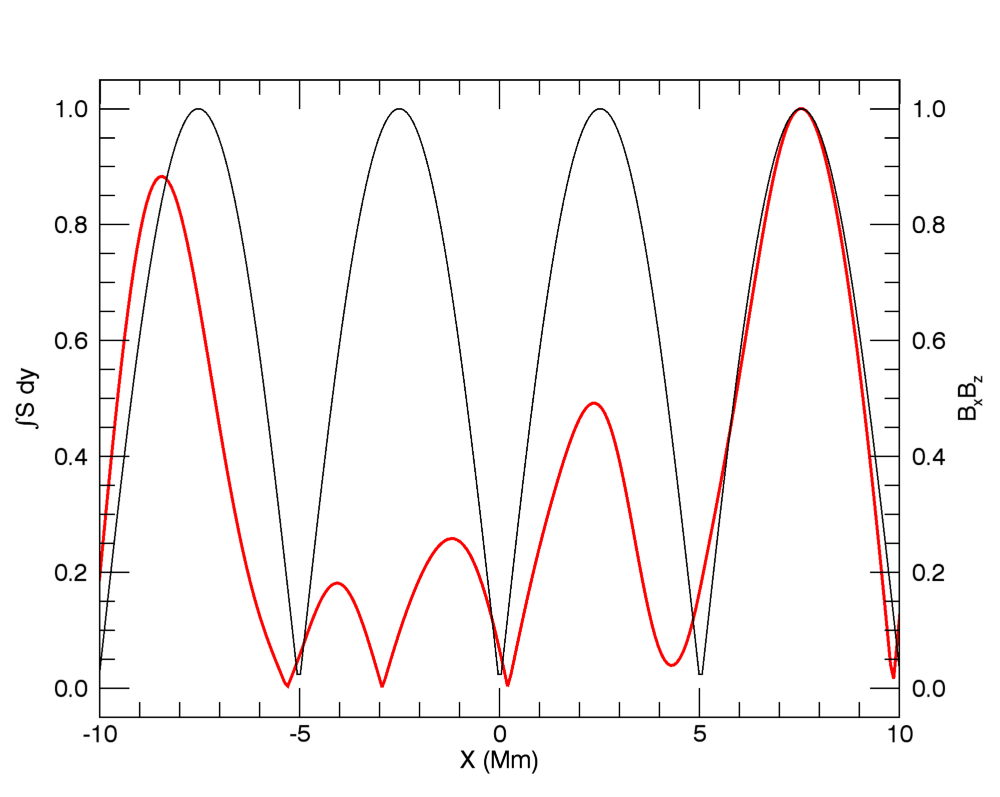}
  \caption{\emph{Red line}: Magnitude of the time-averaged Poynting flux through the driven boundary as a function of $x$ for the ideal T2 simulation. \emph{Black line}: The initial value of $B_xB_z(x)$. For both lines, we have normalised by the maximum values.}
  \label{PF_profile}
\end{figure}

In Fig. \ref{PF_profile}, we show the magnitude of the time-averaged Poynting flux (red line) for $t < 500$ s. To reduce the effects of random variance, we have integrated along the length of the coronal arcade (-10 Mm $\le y \le$ 10 Mm). For comparison, we also show the initial profile of $|B_xB_z|$. By comparing the two lines, we see that, as expected, the Poynting flux is small where $|B_xB_z|$ is close to 0. 

Over long time periods (reducing the effects of variance in the velocity profiles), the action of the imposed velocity driver on the equilibrium field profile will not directly lead to a net change in the total energy within the domain. This is because initially, the integral of $B_xB_z$ over the lower boundary is zero, and the $v_x$ and $v_y$ terms in equation \ref{Poynt_flux_exp} will also have a mean contribution of zero. Instead, the energy injection arises from the effects of the velocity profile on the perturbed magnetic field. For example, a positive $v_x$ component, will induce a negative component in the perturbed $B_x$ and thus, assuming that the velocity does not instantaneously change direction, will lead to a net flux of energy into the domain.

\begin{figure}[h]
  \centering
  \includegraphics[width=0.47\textwidth]{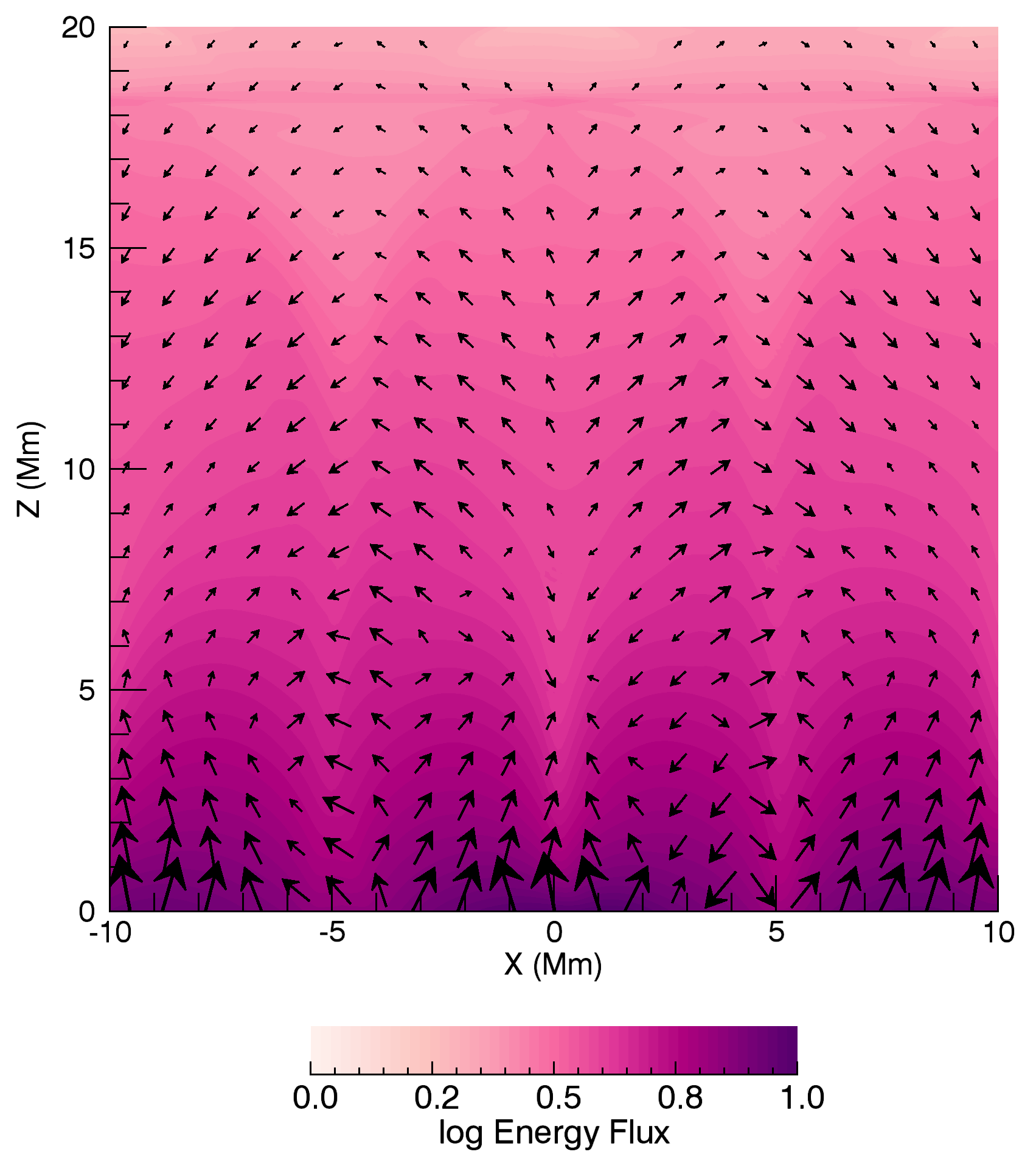}
  \caption{\emph{Vectors:} Energy flux averaged over the length of the arcade and the duration of the ideal T2 simulation. \emph{Contours:} Magnitude of the energy flux. For clarity, we show the logarithm of this quantity.}
  \label{Energy_flux}
\end{figure}

Although the injected Poynting flux is spatially non-uniform, this does not lead (directly) to some field lines becoming consistently more energetic. This is because energy is continuously redistributed throughout the domain during the simulation. In order to track this redistribution, we calculate the energy flux vector, $\vec{F}$, defined by
\begin{equation}
\vec{F} = \frac{\rho v^2 \vec{v}}{2} + \frac{\gamma P \vec{v}}{\gamma -1} + \frac{\vec{E} \times \vec{B}}{\mu}.
\end{equation}
Here, the first term is the kinetic energy flux, the second term is the enthalpy flux and the final term is the Poynting flux.
We then average this quantity in time over the duration of the simulation and spatially along the $y$ axis (the length of the arcade). 

The resultant vector field is displayed in Fig. \ref{Energy_flux}. The contours show the magnitude of the energy flux and the vectors show its projection onto the $x$-$z$-plane. The magnitude of the vector field is much larger close to $z = 0$ Mm than at higher altitudes. Hence, for clarity, we show the logarithm of the averaged energy flux in the contour plot. The length of the arrows also reflects the logarithm of the magnitude of the projected vectors. Everywhere within the domain, the flux of energy is dominated by the Poynting flux. The enthalpy flux is non-neglibigble (but still smaller than the Poynting flux) at large $z$, where the plasma-$\beta$ is higher. The kinetic energy flux is very small in comparison throughout the arcade. 

Since the source of additional energy is the lower $z$ boundary, the rapid decrease in the magnitude of the energy flux with height indicates that most of the additional energy is confined to low altitudes within the arcade. Indeed, it is stored almost exclusively in the perturbed field close to $z = 0$ Mm (see Fig. \ref{Isosurfaces}) in all simulations. As we shall see (Sect. \ref{section:non-ideal}), this will ensure that the majority of irreversible plasma heating occurs close to magnetic foot points. 

Whilst we forced the mean kinetic energy of the driver to be approximately constant across the T1, T2 and T3 simulations, the Poynting flux need not be the same. In particular, since flows are more long-lasting in the T3 simulations, larger deviations from the initial magnetic field are able to form. As the injected energy arises from the effect of the velocity profile on these deviations, a greater Poynting flux into the domain is obtained for the longer time scale cases.

\begin{figure}[h]
  \centering
  \includegraphics[width=0.49\textwidth]{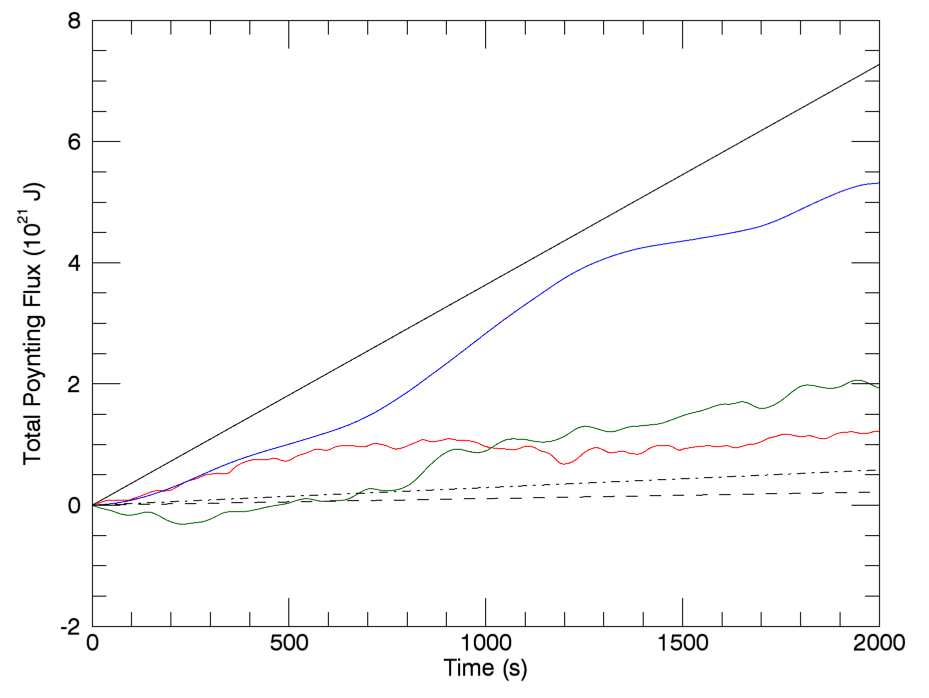}
  \caption{Total energy injected by driver for T1 (red), T2 (green) and T3 (blue) simulations). For comparison, we also show approximate energy requirements in an active region (solid black), Quiet Sun (dashed black) and coronal hole (dot-dashed black).}
  \label{PF_time}
\end{figure}

\begin{figure*}[h]
  \centering
  \includegraphics[width=\textwidth]{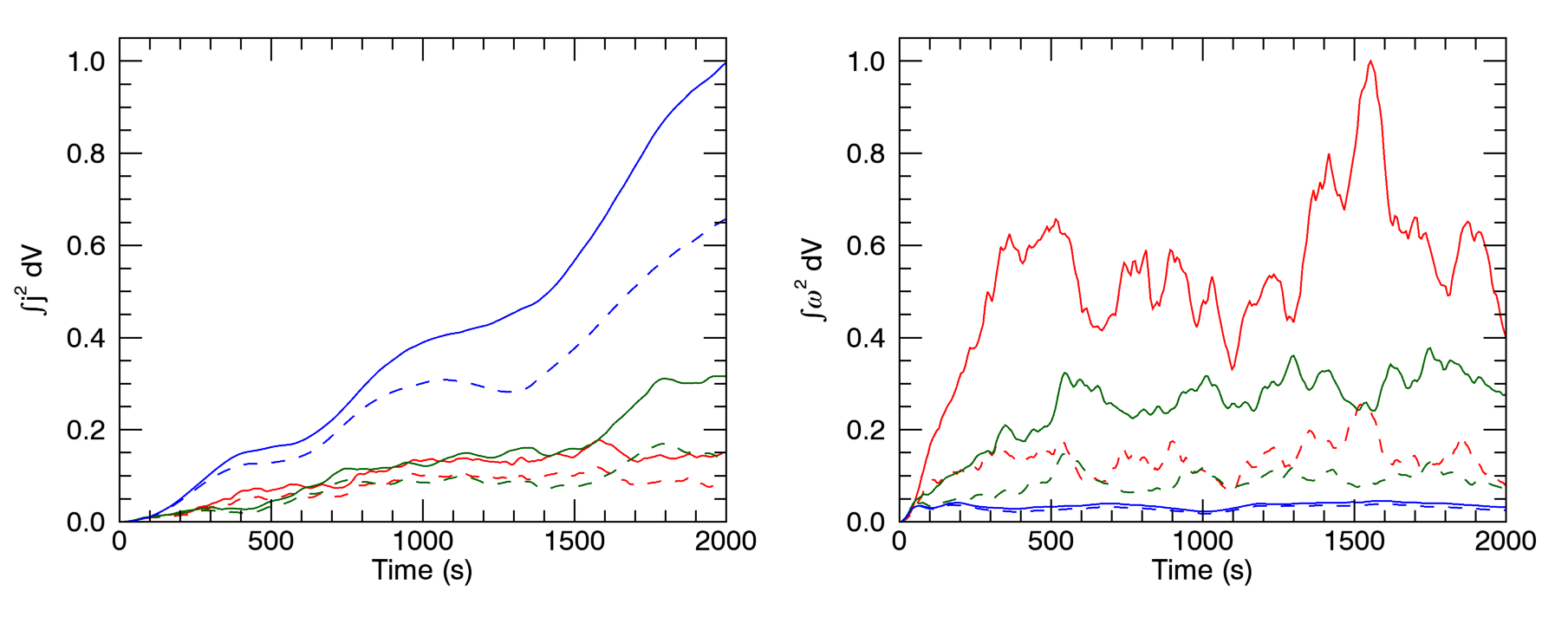}
  \caption{\emph{Left:} Volume integral of the square of the current density throughout the T1 (red), T2 (green) and T3 (blue) simulations. We show results for ideal (solid lines) and resistive (dashed) experiments. \emph{Right:} Volume integral of the square of the vorticity for the T1 (red), T2 (green) and T3 (blue) simulation. We show results for ideal (solid lines) and viscous (dashed) experiments. In each panel, we have normalised the curves using the maximum values obtained during the simulations.}
  \label{Vort_Cur_Growth}
\end{figure*}

In Fig. \ref{PF_time}, we show the total, cumulative Poynting flux as a function of time for the ideal T1 (red), T2 (green) and T3 (blue) simulations. As expected, there is an increase in Poynting flux for longer driving time scales. We note that due to the random nature of the driver, the energy within the domain does not necessarily increase monotonically. Indeed, for $t \lesssim 250$ s, the driver actively removes energy from the domain in the T2 simulation (green curve).  As a result of the general behaviour observed in Fig. \ref{PF_time}, more energy is available to be dissipated for the long time scale drivers. As such, if the heating efficiency is constant across resistive versions of the simulations, then higher plasma temperatures will form in the T3 simulation. 

In Fig. \ref{PF_time}, we also include approximate energy requirements for typical active region (solid black line) and Quiet Sun (dashed black line) conditions \citep{Withbroe1977}. Although not directly applicable to this closed field topology, we also show the required energy budget for coronal holes (dot-dashed line). We note that the magnetic field strength at the base of our simulations is most relevant to active region environments. Whilst slightly below the requirements for active region heating, the long time scale driver (blue line) injects energy at a rate that is comparable to the solid black line. This is despite the relatively low velocities imposed at the lower boundary. On the other hand, the short time scale driving observed in the T1 and T2 simulations, does not provide sufficient energy to balance active region losses. However, even in these two cases, the injected energy could heat the Quiet Sun if the dissipation rate was sufficiently large. This is in contrast to results investigating sinusoidal wave drivers \citep[such as][]{Prok2019, Howson2020} which show that only high amplitude wave drivers will inject enough energy in low-dissipation regimes. The key difference in these simulations is that the driver will introduce complex tangling into the field. This is not the case with a simple, periodic wave driver, which will remove energy from the perturbed field for approximately half of the driving time (with the exception of resonant field lines).

\subsection{Currents and vorticities}
In this section, we investigate the rate of small scale formation in the magnetic and velocity fields. For the former, we use the current density. The square of this quantity is proportional to the Ohmic heating in a resistive regime. For the latter, we use the vorticity. This is not directly related to the complex viscous heating terms but is generally a good proxy for the magnitude of heating observed in a viscous regime.

\begin{figure*}[h]
  \centering
  \includegraphics[width=0.95\textwidth]{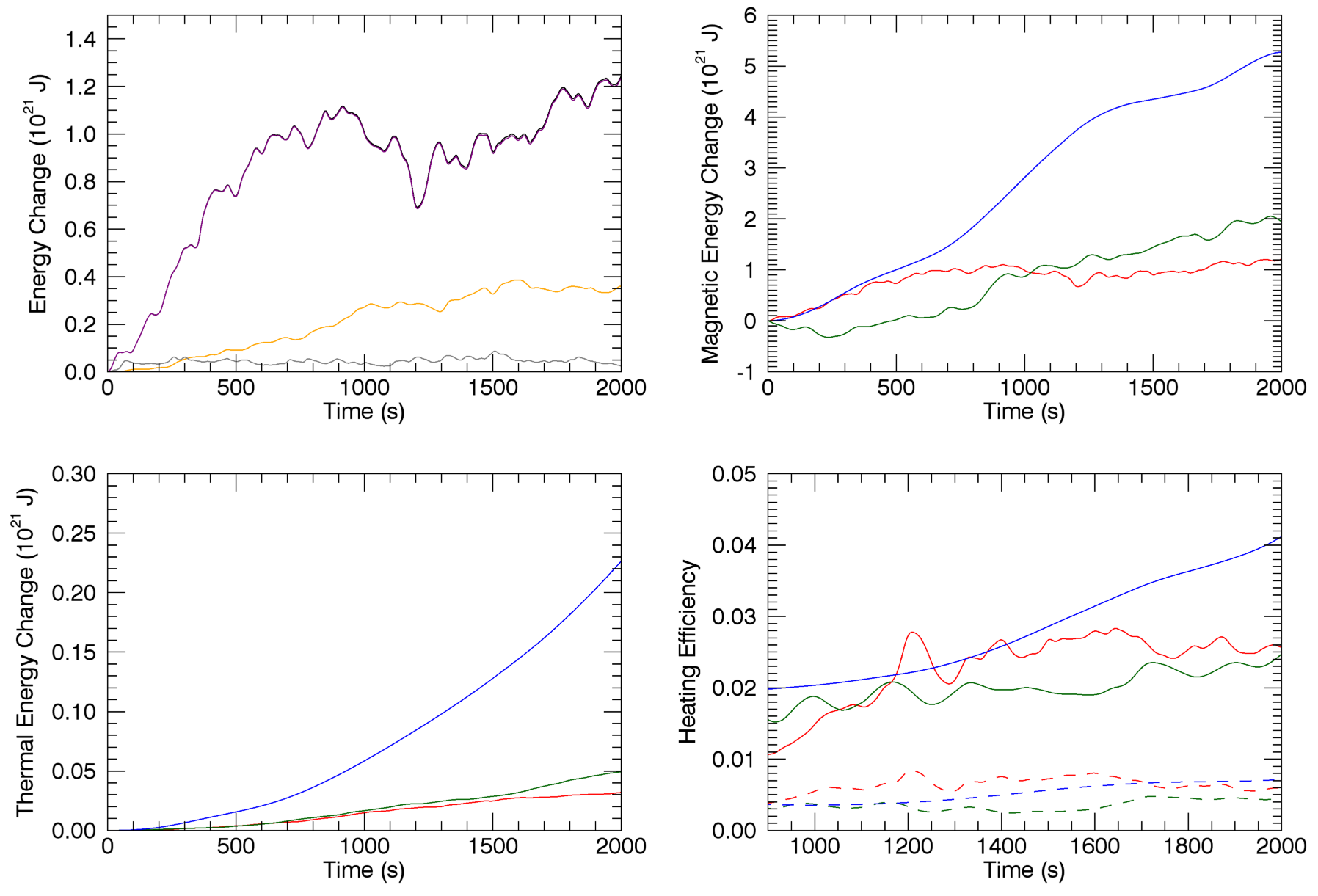}
  \caption{\emph{Upper left:} Energy increase in the T1, ideal simulation. We show each component of the energy; magnetic (purple), thermal (orange) and kinetic (grey). In order to make the latter two lines visible, we have multiplied by factors of 50 and 500, respectively. The total energy change is shown in black and is almost identical to the purple line. \emph{Upper right:} Increase in volume integrated magnetic energy for resistive T1 (red), T2 (green) and T3 (blue) simulations. \emph{Lower left:} Increase in volume integrated thermal energy for resistive T1 (solid red), T2 (solid green) and T3 (solid blue) simulations. \emph{Lower right:} Heating efficiency for resistive T1 (red), T2 (green) and T3 (blue) simulations. This is calculated as the ratio between the increase in thermal energy and the total energy increase.}
  \label{Heating_info}
\end{figure*}

In the following, we consider a set of resistive simulations with magnetic Reynolds numbers of approximately $10^4$ and viscous simulations with fluid Reynolds number of approximately $10^3$. Although these values are much larger than might be expected within the solar corona, they ensure that energy is dissipated by the user-imposed resistivity and viscosity and not by numerical effects. Substantially larger Reynolds numbers are difficult to obtain in these large scale three dimensional MHD simulations due to computational constraints. 

For the resistive simulations we implement a resistivity, $\eta$, that is uniform everywhere apart from close to the $z=0$ boundary, where it is set to zero. In particular, it is zero for $z < 1$ Mm and a constant value, $\eta_0$ (to give the required magnetic Reynolds number), otherwise. This profile reduces the slippage of magnetic field lines through the velocity field imposed at the lower $z$ boundary. It also means that there is no Ohmic heating for $z < 1$ Mm. The viscous simulations, on the other hand, use a uniform viscosity profile. The driving imposed in the resistive, viscous and ideal forms of each simulation (T1, T2 and T3) were identical. We also note that these simulations do not consider the loss of energy through thermal conduction or optically thin radiation or by the exchange of mass between the corona and lower layers of the solar atmosphere. As a result, the plasma temperatures obtained (see Sect. \ref{section:non-ideal}) are higher than would be expected for comparable energy release in the real corona.

In Fig. \ref{Vort_Cur_Growth}, we display the time evolution of the current density (left panel) and vorticity (right panel). In order to reduce the effects of the random variance associated with the velocity drivers, we display the volume integral of the magnitude of these vector quantities. In both panels, we show the time evolution for the T1 (red), T2 (green) and T3 (blue) simulations. The solid lines represent ideal experiments and the dashed lines show the resistive (left panel) and viscous (right panel) simulations. In the case of the current density (left panel), we also show the results for resistive simulations (dashed lines) and for the vorticity (right panel), we show the results of viscous simulations (dashed lines). 

Beginning with the left-hand panel, we see that the currents are larger in the simulations with longer time scale driving (blue lines). This is largely due to the increased energy input in these simulations, and in particular, the increased magnetic energy. Since the perturbed magnetic field is not potential, it is associated with currents and, in general, the larger the field strength, the larger the currents. Meanwhile, the reduction of the currents in the resistive simulations are simply a result of the diffusion of strong gradients in the field by the magnetic diffusivity. 

For the vorticity, on the other hand, we see the opposite behaviour. In the right-hand panel of Fig. \ref{Vort_Cur_Growth}, we observe that the short time scale drivers (red and green lines) induce larger vorticities. This can be understood by considering the manner in which driving time scales map to velocity length scales within the coronal plasma. A high frequency, horizontal, sinusoidal driver imposed on a magnetic foot point will excite a transverse wave that can propagate along field lines. The frequency of the driver will correspond to the wave length of the excited wave. High frequencies will produce short wavelengths and therefore large velocity gradients and, conversely, low frequencies will produce long wavelengths and consequently, small velocity gradients. Of course, in this case, the velocity driving is random, but the same principle applies. The shorter time scale, higher frequency drivers induce large gradients in the velocity field and therefore increase the volume integrated vorticity. 

As with the current densities, we note a drop in the vorticities for the dissipative simulations (dashed lines). Relative to the magnitude of the volume integrated gradients, this drop is more significant in the case of the vorticities (compared to the current densities). This behaviour can be understood by considering the relative Reynolds numbers in the dissipative experiments. In particular, the fluid Reynolds number is an order of magnitude lower in the viscous cases than the magnetic Reynolds number is in the resistive simulations. This means that the relatively larger viscosity is more effective at suppressing gradients in the velocity field, and hence, vorticities are reduced by a greater fraction. Despite this, we note that in all simulations there is significantly more magnetic energy than kinetic energy. This is still the case when only the energy in the perturbed (not background) magnetic field is considered. Consequently, unless viscous effects are many orders of magnitude greater than resistive effects, we expect Ohmic heating to dominate over viscous heating. As such, it will be the size of the currents, and not the vorticities, that is important for energy dissipation and plasma heating.

\subsection{Energy dissipation} \label{section:non-ideal}
The gradients in the magnetic and velocity fields discussed in the previous section are susceptible to dissipation in the non-ideal simulations. In the upper left-hand panel of Fig. \ref{Heating_info}, we show the change in components of the volume integrated energy during the ideal T1 simulation. We show the change in magnetic (purple), internal (orange) and kinetic (grey) energies. In order to make the latter two curves visible, the internal energy has been multiplied by a factor of 50 and the kinetic energy by a factor of 500. It is clear that the change in energy is dominated by the increase in magnetic energy. This is true across all ideal, resistive and viscous simulations. We also show the total energy change (black) but this is largely obscured by the purple curve. Although we have labelled this simulation as ideal, shock viscosities are included and these contribute small dissipative effects which heat the plasma. Of course, in the equivalent resistive and viscous simulations, the increase in internal energy is larger, however it is still much smaller than the change in magnetic energy.

In the upper right-hand panel of Fig. \ref{Heating_info}, we show the change in magnetic energy for the resistive T1 (red), T2 (green) and T3 (blue) simulations. This perturbed magnetic energy is the source for the majority of the thermal energy release in all experiments. As the change in magnetic energy is almost identical to the total energy change, it is not surprising that the differences between the simulations mirror those observed in the cumulative Poynting flux (Fig. \ref{PF_time}). Again, we see larger energy increases for simulations with longer driving time scales. 

In the lower left-hand panel of Fig. \ref{Heating_info}, we show the change in thermal energy for the resistive T1 (red), T2 (green) and T3 (blue) simulations. In each case, this reflects the irreversible dissipation of energy that occurs during the simulation and is dominated by Ohmic heating. We note that for the longest time scale driving (blue), the volume and time integrated heating is almost an order of magnitude larger than for the shortest time scale driving (red). This is despite the imposed drivers having approximately the same kinetic energy.

We note that, since there are no energy loss mechanisms included in these simulations, in a statistically steady state, the volume integrated thermal energy should increase approximately linearly. We see that by the simulation end-time, this is the case for the T1 (red) and T2 (green) simulations but not in the T3 (blue) case. Indeed, in this latter simulation, the rate of energy release is still increasing rapidly at $t = 2000$ s. Unfortunately, computational constraints prevent this numerical experiment running for sufficient time to reach a statistically steady state. Despite this, we note that the long term energy release will be even larger for the T3 simulation and thus the heating rate will dominate over the shorter time scale driving by an even greater amount.

In the lower right-hand panel of Fig. \ref{Heating_info}, we show the heating efficiency for the resistive (solid lines) and viscous (dashed lines) forms of the T1 (red), T2 (green) and T3 (blue) simulations. This is calculated as the ratio between the change in internal energy and the change in total energy. In other words, we determine how much of the injected energy is dissipated as heat. We note that there is a small amount of energy lost through the damping layer and this is not accounted for in this calculation. However, since most of the injected energy is confined to low values of $z$, this has little effect on the calculated efficiencies. Further, we highlight that the first 900 s of the simulation have been omitted to exclude times when the energy has been removed by the driver in the T2 simulation. As expected, due to the relative sizes of the injected magnetic and kinetic energies, the heating efficiency is larger in the resistive simulations than in the viscous cases. However, even in the simulation with the greatest heating efficiency (resistive T3), the amount of heating is relatively low (< 5\% of energy is converted to heat). We note that such a low efficiency could not be maintained over much longer time periods. As such, if the energy injection rate does not decrease, additional energy must be released through an increased heating rate or via larger scale events such as flares and/or coronal mass ejections.

\begin{figure}[h]
  \centering
  \includegraphics[width=0.47\textwidth]{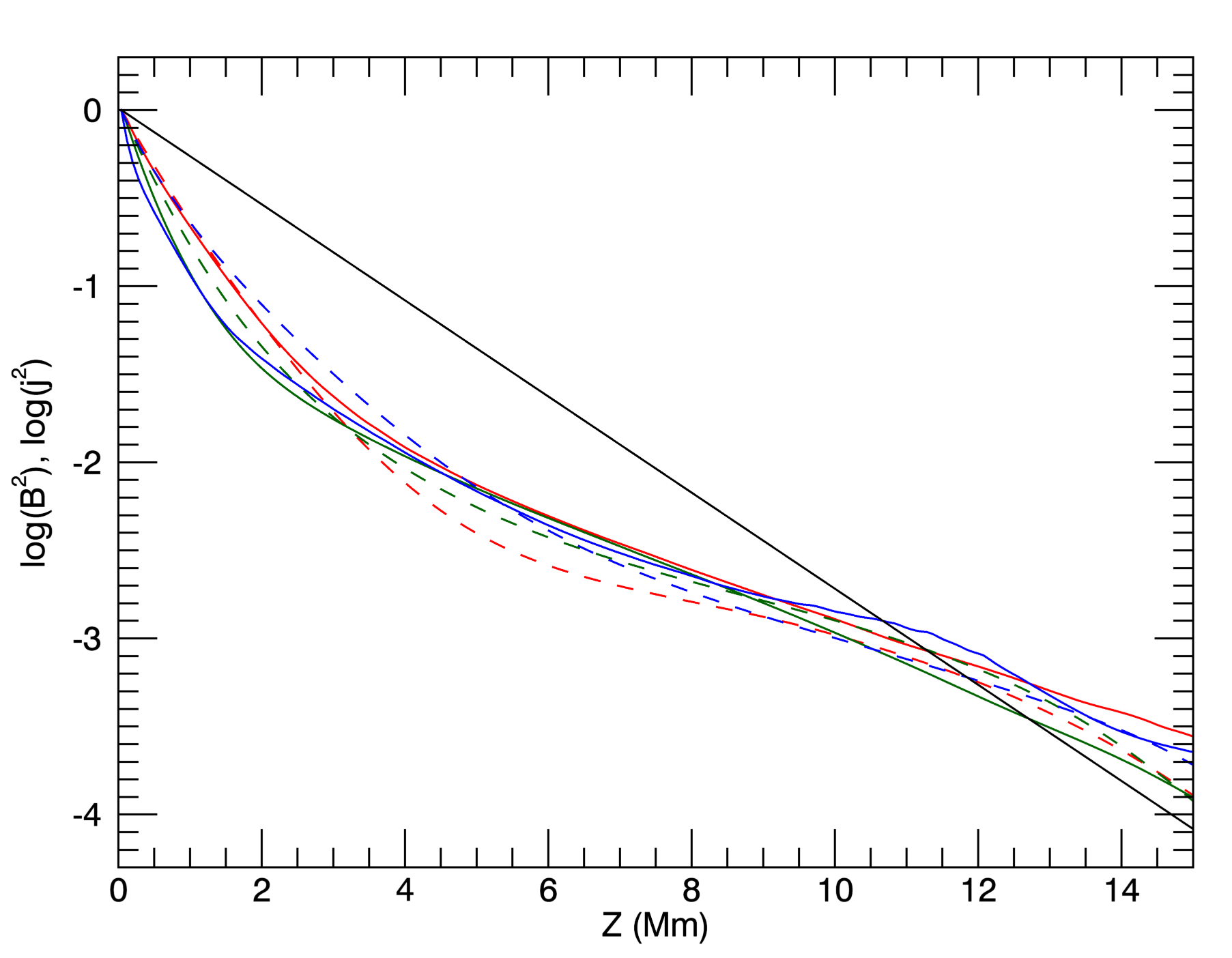}
  \caption{Mean of $j^2$ as a function of height at the end of the T1 (solid red), T2 (solid green) and T3 (solid blue) simulations. We also show the mean of the perturbed magnetic field strength for the T1 (dashed red), T2 (dashed green) and T3 (dashed blue) simulations. The solid black line shows the mean of the initial magnetic field strength as a function of height. For all variables we have normalised by their respective maxima and taken logarithms.}
  \label{j_heights}
\end{figure}

\begin{figure*}[h]
  \centering
  \includegraphics[width=\textwidth]{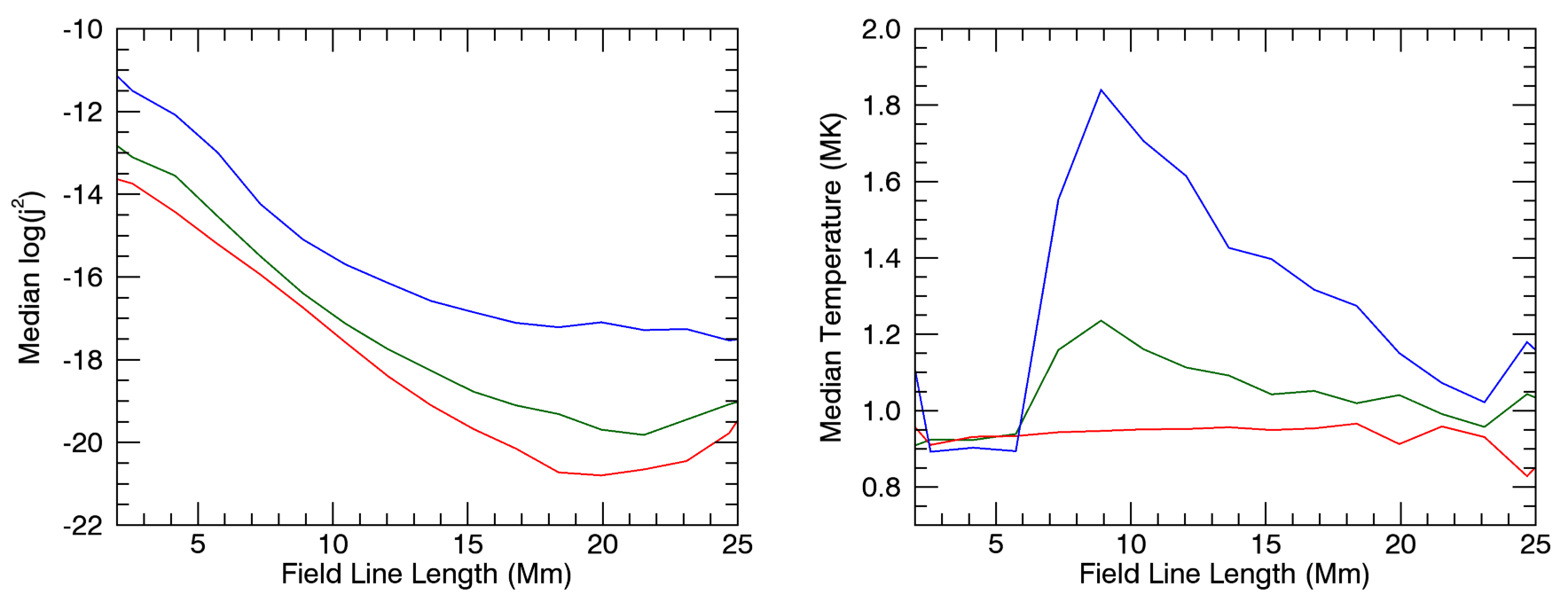}
  \caption{\emph{Left:} Median $\log{j^2}$ along field lines of different lengths at the end of the T1 (red), T2 (green) and T3 (blue) simulations. \emph{Right:} Median temperature along field lines of different lengths at the end of these simulations.}
  \label{fieldline_heating}
\end{figure*}

In Fig. \ref{j_heights}, we show the mean of $j^2$ (solid curves) at different heights within the domain at the end of the T1 (red), T2 (green) and T3 (blue) resistive simulations. To allow for comparison of the profiles between experiments, in each case, we have normalised to the maximum value in the respective simulation. We also plot the mean of the perturbed magnetic field strength (dashed line) at different heights within the same simulations. We have normalised this quantity in the same way as the $j^2$ curves. The solid black line shows the mean strength of the initial magnetic field as a function of height. Again, this is normalised to the maximum value which is obtained at the base of the domain. In this figure, all quantities are plotted logarithmically. 

In these resistive simulations, $j^2$ is proportional to the Ohmic heating rate, except for $z < 1$ Mm, where the resistivity is set to 0. We see that in all cases, the heating is much larger at low altitudes within the domain (akin to foot point heating) and also scales with height in a similar manner in all three simulations. Further, this scaling in the heating rate corresponds to the fall-off in the perturbed magnetic field strength (dashed lines). It does not scale directly with the equilibrium magnetic field (black line). This is understandable as the background field is current-free and thus will not contribute to heating the plasma. However, since the total field strength is dominated by the initial field, it is actually a poor predictor of the local heating rate. This will always be the case if the coronal field is comprised of a dominant potential field and only a weak non-potential component.

In Fig. \ref{fieldline_heating}, we investigate the heating rate and temperature increase on field lines of different lengths. At the end of the three resistive simulations, we traced $10\,000$ magnetic field lines with foot points uniformly distributed on the lower $z$ boundary. Along each field line, we calculated the median value of $j^2$ and of the temperature. The field lines were then binned according to their length and the median value of $j^2$ (left panel) and the temperature (right panel) in each bin are displayed in Fig. \ref{fieldline_heating}.

Beginning with the left-hand panel, we see that the shortest field lines have much larger average values of $j^2$ (note the logarithmic scale). This is simply because the entire length of the field line is at low altitudes and thus confined within the high current region. As such, short field lines can be heated much more efficiently than long field lines. This is borne out in the right hand panel of Fig. \ref{fieldline_heating} which shows that, in general, the median temperature decreases as the field line length increases. Of course, this is not the case for the shortest field lines as these have a large proportion of their length within the zero resistivity volume at the base of the domain. In agreement with our previous analysis, we see that the shorter time scale driving simulations (red and green curves) exhibit lower currents and therefore temperatures, on all field line lengths compared to the long time scale driving case (blue curve).

\begin{figure}[h]
  \centering
  \includegraphics[width=0.47\textwidth]{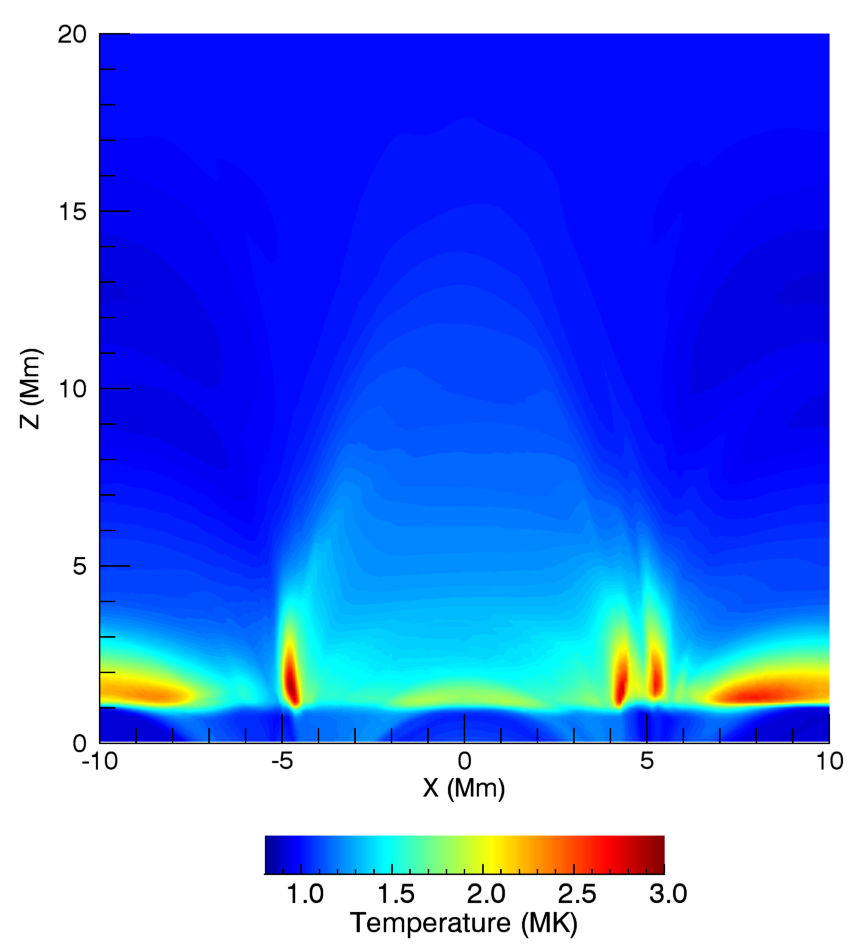}
  \caption{Median temperature along the arcade at $t = 2000$ s in the resistive T2 simulation.}
  \label{Temp_eta_0.1}
\end{figure}

In Fig. \ref{Temp_eta_0.1}, we display the resultant temperature profile at $t = 2000$ s in the resistive T2 simulation. Specifically, in order to reduce the effects of the random variance in the imposed driver, we show the temperature averaged (median) along the length of the coronal arcade. Whilst the magnitude of the heating is different in the T1 and T3 simulations, the locations of the highest temperature regions are (approximately) the same. We also note that the values of the temperature displayed in Fig. \ref{Temp_eta_0.1} are somewhat arbitrary due to the lack of energy loss mechanisms considered in these simulations.

We immediately notice the effects of the zero resistivity region close to the lower boundary which prohibits Ohmic heating for $z < 1$ Mm. The largest increase in temperature occurs just above this region, where the currents are still large and $\eta \ne 0$. High temperatures form in low lying coronal loops and also near the base of the separatrix surfaces between the arcades (see discussion of Fig. \ref{Isosurfaces}). From here, gas pressure forces are able to redistribute some high temperature plasma along magnetic structures. This can be seen by the presence of moderately high temperature material at $z<1$ Mm. This hotter plasma is only located at the base of arcades connected to the resistive volume and is absent from field lines that are contained solely within the $\eta = 0$ region (e.g. -1 Mm $ < x < 1$ Mm). We note that if conduction was to be included in these simulations, we would expect the thermal energy to spread along magnetic field lines more efficiently and thus reveal more of the arcade structure in the temperature profile.

\begin{figure}[h]
  \centering
  \includegraphics[width=0.47\textwidth]{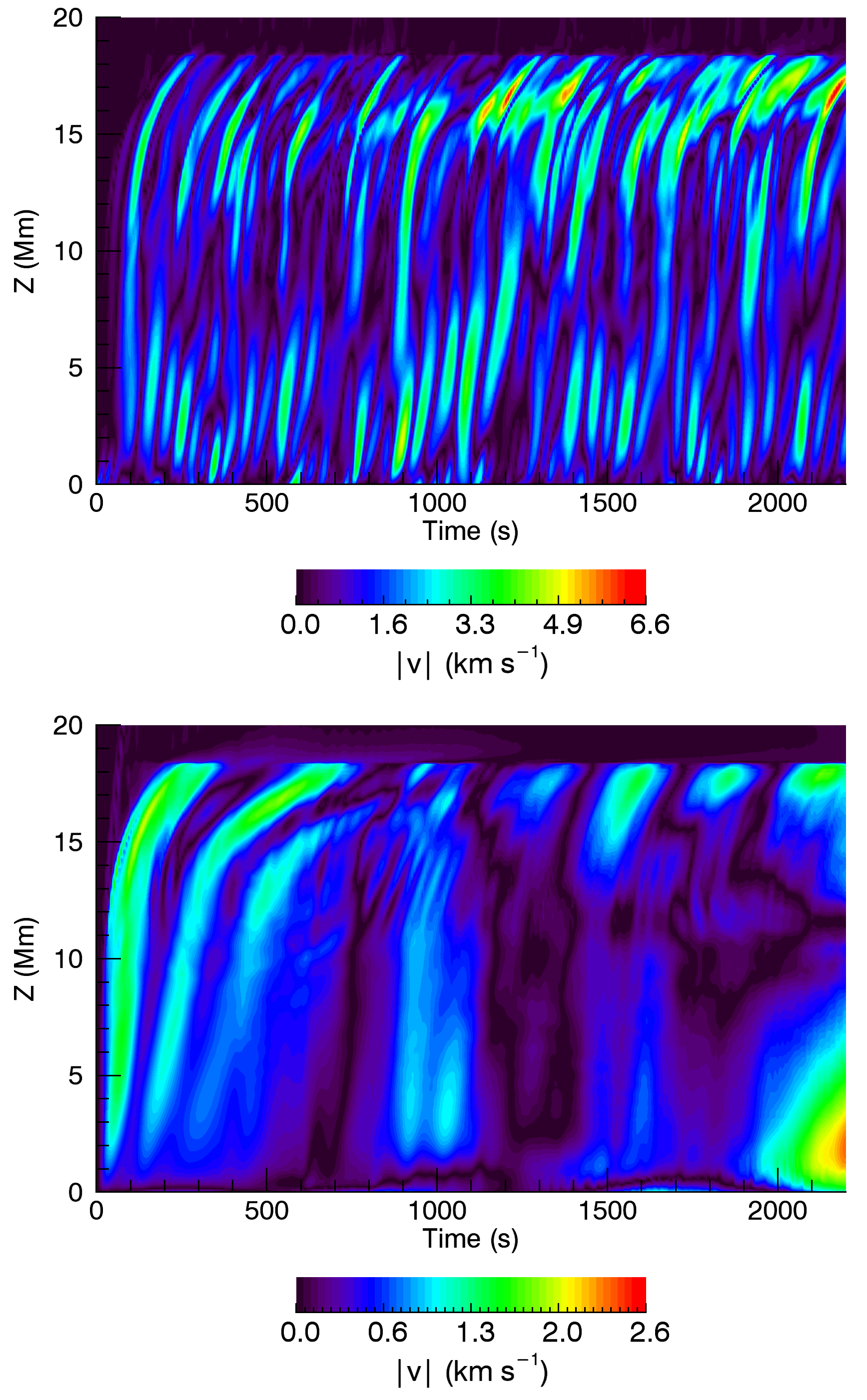}
  \caption{Evolution of $|v_x|$ along the line $x= y=0$ for the ideal T1 (upper) and T3 (lower) simulations. We note that the colour bar for each panel is different.}
  \label{vel_time}
\end{figure}

\subsection{Velocity and kinetic energy} \label{section:vel_ke}
A key difference between the T1, T2 and T3 simulations is apparent in the velocity fields generated by the imposed driving. Although these are energetically insignificant in comparison to the magnetic fields, they are much more easily observed in the corona. The driving in the T1 simulation is associated with short time scales throughout the simulation domain (and not only at the lower boundary). In particular, wave-like behaviour is apparent in the T1 velocity field whereas, only slow, gradual evolution is observed in the T3 case.

\begin{figure}[h]
  \centering
  \includegraphics[width=0.47\textwidth]{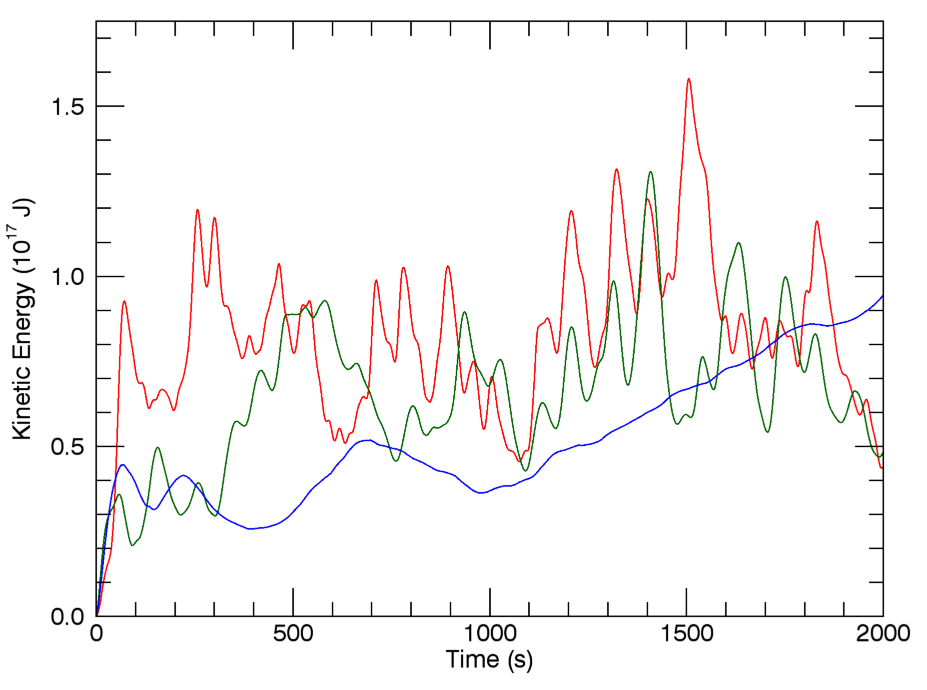}
  \caption{Volume integrated kinetic energy as a function of time for the ideal T1 (red), T2 (green) and T3 (blue) simulations.}
  \label{KE_compare}
\end{figure}

In Fig. \ref{vel_time}, we show the evolution of $|v_x|$ along the line $x = y = 0$ for the ideal T1 (upper panel) and T3 (lower panel) simulations. This corresponds to a vertical line through the centre of the arcade. The corresponding figures for the resistive and viscous simulations are very similar. When observing along the line of sight parallel to the (initially) invariant direction of the magnetic arcade, this component of the velocity would be identified as transverse motions in the plane of the sky. In both panels, the region of low velocity above $z = 18$ Mm, corresponds to the imposed damping layer at the top of the simulation domain. With the exception of this region, the distribution of kinetic energy throughout the domain is much more uniform than the distribution of magnetic (both total and perturbed) energy.

We can immediately observe shorter temporal scales in the T1 simulation when compared to the T3 case. Velocity features have much shorter lifetimes and occur much more frequently. This behaviour generates the enhanced vorticity for the short time scale driving simulations discussed previously. In both panels of Fig. \ref{vel_time}, features can be seen to propagate from low heights upwards through the domain. This phenomenon is generated by a combination of fast waves propagating across the magnetic field and as a result of phase mixing; perturbations excited at the magnetic foot points take longer to propagate to the apex of longer field lines. The reduction in the Alfv\'en and fast speeds at high altitudes causes the decrease in gradient observed near the top of both plots. 

Whilst the largest velocities are greater in the T1 simulation, the mean magnitude of flows throughout the computational volume does not change drastically for different characteristic driving time scales. To show this, in Fig. \ref{KE_compare}, we show the time-evolution of the volume integrated kinetic energy for the ideal T1 (red), T2 (green) and T3 (blue) simulations. Whilst the T1 simulation does show slightly higher kinetic energies than the T3 case, the difference is not large, and indeed, this behaviour reverses near the end of the simulation run time. It is certainly possible that this difference can simply be explained by the random nature of the drivers and need not be associated with the time scale of the driving. Without conducting a much larger parameter study, the existence of any genuine difference remains unclear. Regardless, the kinetic energy is so small in comparison to the magnetic energy (both initial and perturbed), that any difference here is energetically irrelevant.

We note that the characteristic driving time scale is once again apparent in the volume integrated kinetic energies. In particular, the red and green curves in Fig. \ref{KE_compare} show much greater variation than the long time scale blue curve. However, whilst this quantity may be estimated from observations (for example using Doppler velocities along an appropriate line-of-sight), it drastically underestimates the total energy in the domain. Additionally, it is not even a good  
proxy of the energy available for heating as it does not increase with the energy in the perturbed magnetic field (compare with top-right hand panel of Fig. \ref{Heating_info}).

\section{Discussion and Conclusions} \label{Sec_Dis}
In this article, we have investigated the effects of driving time scales on energy injection and dissipation within a coronal arcade. Whilst we forced the driver power (in terms of its kinetic energy) to be independent of the characteristic time scale, simulations with longer time scale driving were found to have increased energy fluxes into the domain. Larger perturbations to the initial potential field were attained for long time scale driving and, consequently, the average Poynting flux was greater in these cases. The increased energy injection ensured that, in resistive regimes, plasma heating was greater for long time scale driving. In all simulations, the vast majority of injected energy is stored in the perturbed magnetic field, which is largest close to the driven boundary. Only a small fraction of the injected energy is transmitted to high altitudes, and thus, most heating is confined to the magnetic foot points. Additionally, since the perturbed magnetic energy is orders of magnitude larger than the kinetic energy, in dissipative regimes, Ohmic heating will generally be much greater than viscous heating.

In order to examine the general properties of the system, throughout this article, we have selected to focus on volume averaged quantities rather than specific energy release events. The location, magnitude and timing of individual events will be sensitive to the specific velocity profile imposed and may not be generalisable to other similar drivers. Whilst we have not investigated particular instances of energy release, we note that the limited spatial resolution will have an impact on the exact nature of localised heating and on the evolution of the system as a whole. In particular, we do not accurately reproduce the process of magnetic reconnection in intricate current sheets. In higher resolution simulations that are closer to the very low dissipation regime of the solar corona, we may expect heating to be more localised and bursty in nature with an increased contribution from viscous heating due to large reconnection outflows. The relatively small (magnetic) Reynolds numbers we are forced to consider will ensure that magnetic energy is released at early times and in less stressed fields than would be the case for the Sun. This will, initially at least, artificially enhance the observed heating rates and reduce the amount of stress that can build up within the magnetic field. As a result, the time-averaged Poynting flux will decrease and, ultimately, less energy will be available for heating the plasma \citep[e.g.][]{Klimchuk2015}. This is likely to be more significant in the T3 simulations, in which large currents form most readily.

It is important to note that our driver is imposed at the base of the corona and therefore should not be compared directly to photospheric flows. As the transmission of power through the lower layers of the solar atmosphere is so complicated, it is unclear how flows at the solar surface map to driving at the base of the corona. Instead, it may be more reasonable to construct coronal drivers such that they generate dynamics that reflect observations of coronal waves. For example, by imposing velocity profiles that reproduce observed power spectra and/or the amplitudes of oscillations. Indeed in the T1 simulation (AC driving), the current construction generates flows within the domain that are comparable to coronal observations of the order 30 $\text{km s}^{-1} $. Additionally, in previous studies \citep[e.g.][]{Morton2016}, authors have identified increased oscillatory power at low frequencies within the corona. This result is applicable across the parameter space investigated within this article. As such, contrary to the argument above, the driving power should potentially be higher for the long time scale (T3) simulations or lower for the short time scale (T1) simulations. This would further enhance the heating in the T3 simulation above the level generated in the T1 simulation. In other words, the dissipation rates presented here may be too generous to the AC-heating simulations and, in reality, we would expect the DC-heating to dominate by an even greater extent.

In the literature, many proposed wave heating models rely on a pre-defined density profile that generates gradients in the Alfv\'en speed/frequency and allows phase mixing and mode coupling to develop. Whilst no such density variation is included in the initial conditions considered here, variations in the field line length and magnetic field strength still allow phase mixing to proceed. Despite this, it is likely that wave heating is less efficient in the absence of a non-uniform density profile. Furthermore, in reality, enhanced wave heating may develop if heating events lead to the localised evaporation of dense plasma into the corona. As such, without a full thermodynamic and gravitational treatment of the coupled solar atmosphere, it remains possible that we underestimate the potential for wave heating in this study.

The potential for the driver to add complexity to the background magnetic field regardless of the characteristic time scale of the driver is a departure from many AC heating studies. Frequently, a sinusoidal driver is imposed at the foot points of magnetic field lines and, in such cases, only periodic fluctuations about the background field are induced. In the current study, however, even for short time scale driving, the field is stressed in a manner that is more typical of DC studies. On account of this, the injected Poynting flux is relatively large given the small amplitude flows considered. In comparison, previous studies \citep[e.g.][]{Howson2020} have found wave drivers to inject insufficient Poynting flux despite much larger amplitude oscillations. As such, regardless of the relevant time scale of the velocity driving, the injection of magnetic field complexity into the corona is an important consideration for all heating studies.  


\vspace{1cm}

{\emph{Acknowledgements.}} The research leading to these results has received funding from the UK Science and Technology Facilities Council (consolidated grant ST/N000609/1), the European Union Horizon 2020 research and innovation programme (grant agreement No. 647214). IDM received funding from the Research Council of Norway through its Centres of Excellence scheme, project number 262622. This work used the DiRAC Data Analytic system at the University of Cambridge, operated by the University of Cambridge High Performance Computing Service on behalf of the STFC DiRAC HPC Facility (www.dirac.ac.uk). This equipment was funded by BIS National E-infrastructure capital grant (ST/K001590/1), STFC capital grants ST/H008861/1 and ST/H00887X/1, and STFC DiRAC Operations grant ST/K00333X/1. DiRAC is part of the National e-Infrastructure.

\bibliographystyle{aa}        
\bibliography{Arcade.bib}           

\end{document}